\documentclass[3p,10pt,sort&compress, colorlinks,linkcolor=blue,square, citecolor=blue]{elsarticlex}

\usepackage[utf8]{inputenc}

\usepackage{amssymb}
\usepackage{amsmath,mathrsfs}
\usepackage{color}

\usepackage{extarrows}

\usepackage{verbatim}

\usepackage{lineno}

\usepackage{siunitx}

\usepackage{stfloats}

\usepackage{bm}

\usepackage{graphicx}

\usepackage{caption}
\usepackage{subcaption}

\usepackage{multicol}

\usepackage{multirow}

\usepackage{mhchem}

\usepackage[toc,page]{appendix}

\usepackage{bm}

\usepackage{float}

\usepackage{threeparttable}

\usepackage{sectsty}
\usepackage[toc,page]{appendix}
\usepackage[colorlinks,linkcolor=blue,anchorcolor=green,citecolor=blue]{hyperref}

\newcommand{\figref}[1]{Fig.~\ref{#1}}
\newcommand{\subfigref}[2]{Fig.~\ref{#1}\ce{#2}}
\newcommand{\subsfigref}[3]{Figs.~\ref{#1}\ce{#2} and \ref{#1}\ce{#3}}
\newcommand{\subssfigref}[3]{Figs.~\ref{#1}\ce{#2}-\ref{#1}\ce{#3}}
\newcommand{\tabref}[1]{Table.~\ref{#1}}
\renewcommand{\eqref}[1]{Eq.~$($\ref{#1}$)$}

\newcommand*{\etal}{\textit{et al.}~}

\begin{document}
	
	
	\begin{frontmatter}
    
        \title{Thermal and Microstructural Simulations of Photonic Sintering of Oxide Ceramics: A Two-Scale Scheme}
        \author[els]{Junlong Ma}
        
        \author[els]{Yangyiwei Yang\corref{cor1}}
        \ead{yangyiwei.yang@mfm.tu-darmstadt.de}
       
  	\author[alt]{Julian N. Ebert}
        
        \author[alt]{Wolfgang Rheinheimer}
            
        \author[els]{Bai-Xiang Xu\corref{cor1}}
        \ead{xu@mfm.tu-darmstadt.de}

        \address[els]{Mechanics of Functional Materials Division, Institute of Materials Science, Technische Universit\"at Darmstadt, Darmstadt 64287, Germany}
	
        \address[alt]{Institut für Keramische Materialien und Technologien (IKMT), Universit\"at Stuttgart, Stuttgart 70569, Germany}
			
			\cortext[cor1]{Corresponding author}
		
     \begin{abstract}
     
Photonic sintering (PS) offers an ultra-fast, contact-free alternative to conventional sintering and has demonstrated its potential for enhancing the sinterability of acceptor-doped barium zirconate (BZY) ceramics. However, a central challenge in the PS process lies in achieving precise control over thermal self-stabilization in the presence of complex microstructural effects arising from photonic-ray--induced thermal profiles. To elucidate the interplay among thermal fields, microstructural evolution, and PS process parameters, this study establishes a two-scale, non-isothermal simulation framework. The framework integrates macroscopic heat-transfer simulations, incorporating effective heat conduction and photonic-ray--induced volumetric heating in the porous media, with microscopic non-isothermal phase-field sintering simulations that resolve microstructure evolution under local thermal profile. Scale bridging is achieved through a temperature field transferring and mapping that satisfies Hill-Mandel condition between the macroscopic and microscopic simulations, while maintaining synchronization between their asynchronous time-stepping schemes. After calibrating model parameters against experimental measurements, the framework successfully reproduces the experimentally observed porosity inhomogeneity along the sample depth. The influence of enhanced localized mass transport is further examined through a parametric investigation of surface and grain boundary diffusivities. Overall, the proposed framework demonstrates its feasibility and physical interpretability in establishing process-microstructure relationships for the scalable fabrication of high-performance protonic ceramics.

     \end{abstract}
		
		
     \end{frontmatter}
\def\smco{\ce{SmCo}\text{-1:7}}

\newcommand*{\diff}{\mathop{}\!\mathrm{d}}
\newcommand*{\Diff}{\mathop{}\!\mathrm{D}}

\newcommand*{\pd}[2]{\mathop{}\!\frac{\partial #1}{\partial #2}}
\newcommand*{\varid}[2]{\mathop{}\!\frac{\delta #1}{\delta #2}}
\newcommand*{\fed}[1]{f_\mathrm{#1}}


\def\fedf{\mathscr{F}}


\def\Rsq{\mathrm{R}^2}
\def\mse{\mathrm{MSE}}

\newcommand*{\E}[1]{\mathop{}\!\times 10^{#1}}

\def\deg{^{\circ}}
\def\CI{\text{CI}_{95\%}}

\def\poro{\varphi}
\def\dens{\theta}
\def\mac{\text{mac}}
\def\mic{\text{mic}}
\def\at{\text{at}}

\def\Kten{\mathbf{K}}
\def\Mten{\mathbf{M}}
\def\flux{\mathbf{j}}
\def\Dten{\mathbf{D}}

\def\tmac{t_\text{mac}}
\def\tmacend{t_\text{mac}'}
\def\tmic{t_\text{mic}}
\def\tmicend{t_\text{mic}'}
\def\Dtmac{\Delta_t^\text{mac}}
\def\Dtmic{\Delta_t^\text{mic}}

\def\Keff{K_\mathrm{eff}}

\def\subs{\text{ss}}
\def\gb{\text{gb}}
\def\sf{\text{sf}}
     
\section{Introduction}

Acceptor-doped barium zirconate \ce{BaZr_{1-x}Y_{x}O_{3-\delta}} (BZY) is a promising protonic ceramic that has been extensively investigated for applications in energy conversion and storage devices \cite{yamaguchi2024, park2015enhanced, hossainReviewExperimentalTheoretical2021}. However, limited bulk diffusion and sluggish grain boundary migration demand high sintering temperatures (1600--1700~$^\circ$C) and extended annealing times (exceeding 24~h) to achieve sufficient densification and grain growth, particularly at high Zr contents \cite{loureiro2019, peng2010bazr08y02o3d, sun2011}. This poor sinterability not only causes processing difficulties but also elevates grain boundary resistance, thereby restricting protonic conductivity. Moreover, such high sintering temperatures can induce stoichiometric deviations due to partial barium volatilization and oxide segregation, further degrading the material properties \cite{ebert2024influence, yamazaki2010}.

To improve the sinterability of BZY protonic ceramics, recent studies have explored non-conventional sintering approaches that enhance energy efficiency by replacing conventional indirect heating with direct energy coupling to the sample \cite{rybakovMicrowaveSinteringFundamentals2013, guillonFieldAssistedSintering2014, kruthLasersMaterialsSelective2003, bocanegra-bernalHotIsostaticPressing2004}. These approaches can effectively reduce sintering time and improve processability while minimizing the need for sintering aids, enabling fine control of the chemical composition and further enhanced protonic conductivity \cite{li2020a}. Among these techniques, photonic sintering (PS), also known as blacklight sintering, is particularly attractive because it enables ultra-fast, contact-free densification of samples by direct energy injection via visible and ultraviolet light, as the setup shown in \subfigref{fig:bs_wf_fc}{a}. By exploiting thermal self-stabilization, where optical absorption is balanced by temperature-dependent thermal emission, PS establishes a steady-state temperature, allowing rapid and well-controlled high-temperature sintering \cite{porzBlacklightSinteringCeramics2022, schererBlacklightSinteringBaTiO32023}.
Experimental studies have demonstrated the strong potential of PS in the effective fabrication of various functional ceramics. For example, Porz \etal presented that short-wavelength irradiation enhances heat absorption, while the use of thermally insulating materials such as expandable graphite effectively reduces heat losses \cite{porzBlacklightSinteringCeramics2022}. Furthermore, Scherer \etal optimized pulse durations for BaTiO$_3$ ceramics using xenon flash lamps, enabling rapid densification without structural degradation \cite{schererBlacklightSinteringBaTiO32023}. In parallel, Scheld \etal successfully applied PS to garnet-based solid-state battery electrolytes, where tailored pulse profiles were employed to mitigate thermal stresses and suppress crack formation \cite{scheldBlacklightSinteringGarnetbased2024}.

Although it is conceptually simple, PS faces several challenges, the most critical of which are microstructural effects arising from the photonic-ray–induced thermal profile. As briefly discussed above, thermal self-stabilization arises from a balance between optical absorption and heat emission, predominantly via thermal radiation. However, the relatively stronger thermal emission at the sample surface and edges can eventually lead to thermal inhomogeneity. While the use of thermally insulating substrates can mitigate this effect, residual thermal inhomogeneity may still persist and manifest as a porosity gradient along the thickness direction, with higher porosity typically observed near the bottom of the sample \cite{ebert2025a}. Moreover, precise tuning of the absorbed and emitted power densities is essential for controlling the heating/cooling rates during photonic sintering \cite{porzBlacklightSinteringCeramics2022}. These thermal effects are, in turn, strongly affected by the evolving microstructure, particularly through changes in surface morphology and packing density/porosity. 
As these transient effects and interactions are difficult to probe experimentally, computational simulations provide a powerful and cost-effective way to quantitatively investigate them, thereby enabling microstructure prediction and informed parameter optimization for the PS process.

Simulation of a PS process should be conducted across two distinct length scales. On the macroscale (over 1~\si{mm}), thermal self-stabilization established between heat absorption and radiative emission within the sample. The primary focus on this scale is on modeling and parameterizing the thermal interaction between the incident light ray and the green body. This is typically conducted through heat-transfer simulations incorporating various thermal boundary conditions (BC). Accordingly, several macroscopic finite-element (FE) heat-transfer simulations have been reported to capture the global thermal response during PS process. Porz et al. employed finite element modeling to resolve the temperature field within the sample under varying heat-flow conditions \cite{porzBlacklightSinteringCeramics2022}. Scherer et al. extracted the axial temperature evolution during PS at different stages using a similar approach \cite{scherer2025sintering}. 
On the microscale (around 10-100~\si{\micro m}), individual particles within the green body undergo thermally activated neck formation and coarsening under the local temperature field, leading to changes in local morphology and porosity. Although it has scarcely been applied to PS process, the phase-field method has been widely employed to simulate microstructure evolution during various sintering processes \cite{hotzer2019, yang2020investigation, seiz2024, yang3DNonisothermalPhasefield2019} under coupled physical fields, including thermal, mechanical, and electrical fields \cite{yang2020investigation, yangElastoplasticResidualStressa, wang2024}. This approach is particularly well suited due to its thermodynamic framework that enables straightforward multiphysics coupling and its ability to capture complex interfacial evolution without explicit tracking. 

Despite the availability of mature simulation approaches on the individual length scales, a dedicated two-scale modeling framework that integrates macroscale heat-transfer analysis with microscale phase-field sintering simulations and is specifically tailored to the PS process remains lacking. Integrating the two length scales within a unified simulation framework is essential for PS process, as the thermal profile is governed by macroscale characteristics of the sample, such as its size, thickness, and optical absorptivity and emissivity, while simultaneously determining the local microstructural evolution on the microscale, including neck formation, coarsening, and grain growth. In particular, the thermal-profile–induced inhomogeneities observed experimentally in PS-processed samples \cite{ebert2025a} cannot be recapitulated by conventional isothermal, single-scale microstructural simulations alone. On the other hand, although a fully coupled multiscale model could in principle resolve these coupled effects with high fidelity, its computational cost is too high. These considerations therefore motivate the development of a more efficient yet physically representative two-scale modeling strategy.

For this purpose, based on the fully resolved powder level single-scale simulation \cite{yang3DNonisothermalPhasefield2019}, this work presents a two-scale simulation framework. It couples macroscopic heat-transfer analysis with microscopic phase-field microstructural evolution. With this, it promises an efficient and effective simulation scheme tailored for photonic sintering with a temperature gradient comparatively lower than that of additive manufacturing selective sintering. 
At each transient step, the local thermal profile, notably the temperature and its spatial gradients, resolved by macroscopic heat transfer simulation on the sample, is passed to multiple microscopic domains to perform non-isothermal sintering simulations. Particular emphasis is placed on morphology (mostly pore shape) evolution and porosity changes, as these features play a central role in densification and transport properties. By linking the simulated microstructural mechanisms to experimentally observed pore gradients, this study aims to establish quantitative process–microstructure relationships and to identify optimal sintering parameters, thereby providing guidance for the scalable fabrication of high-performance protonic ceramics.

\section{Method}
\label{methods}
\subsection{Two-scale scheme and scale-bridging}

The simulations performed in this work are based on experimental studies of photonic-sintered acceptor-doped BZY samples reported in Ref.~\cite{ebert2025a}. A blue laser was employed to sinter a cylindrical specimen with a radius of $3000~\si{\micro m}$ and a thickness of $1800~\si{\micro m}$. The incident laser powder density exhibited a uniform distribution with a value of $J_\mathrm{d} = 200~\mathrm{W\,cm^{-2}}$, and the processing sequence consisted of a $30~\si{s}$ heating stage, a $30~\si{s}$ power-holding stage, followed by a $100~\si{s}$ power-decay period. The powder stack has a particle size distribution characterized by a mean diameter of $\bar{d} = 2.879~\si{\micro m}$ and a standard deviation of $\sigma = 3.982~\si{\micro m}$ \cite{ebert2025a}. These experimentally measured parameters were directly adopted in the modeling framework to ensure consistency between the numerical simulations and the experimental conditions.

\begin{figure} [h!]
    \centering
    \includegraphics[width=1\linewidth]{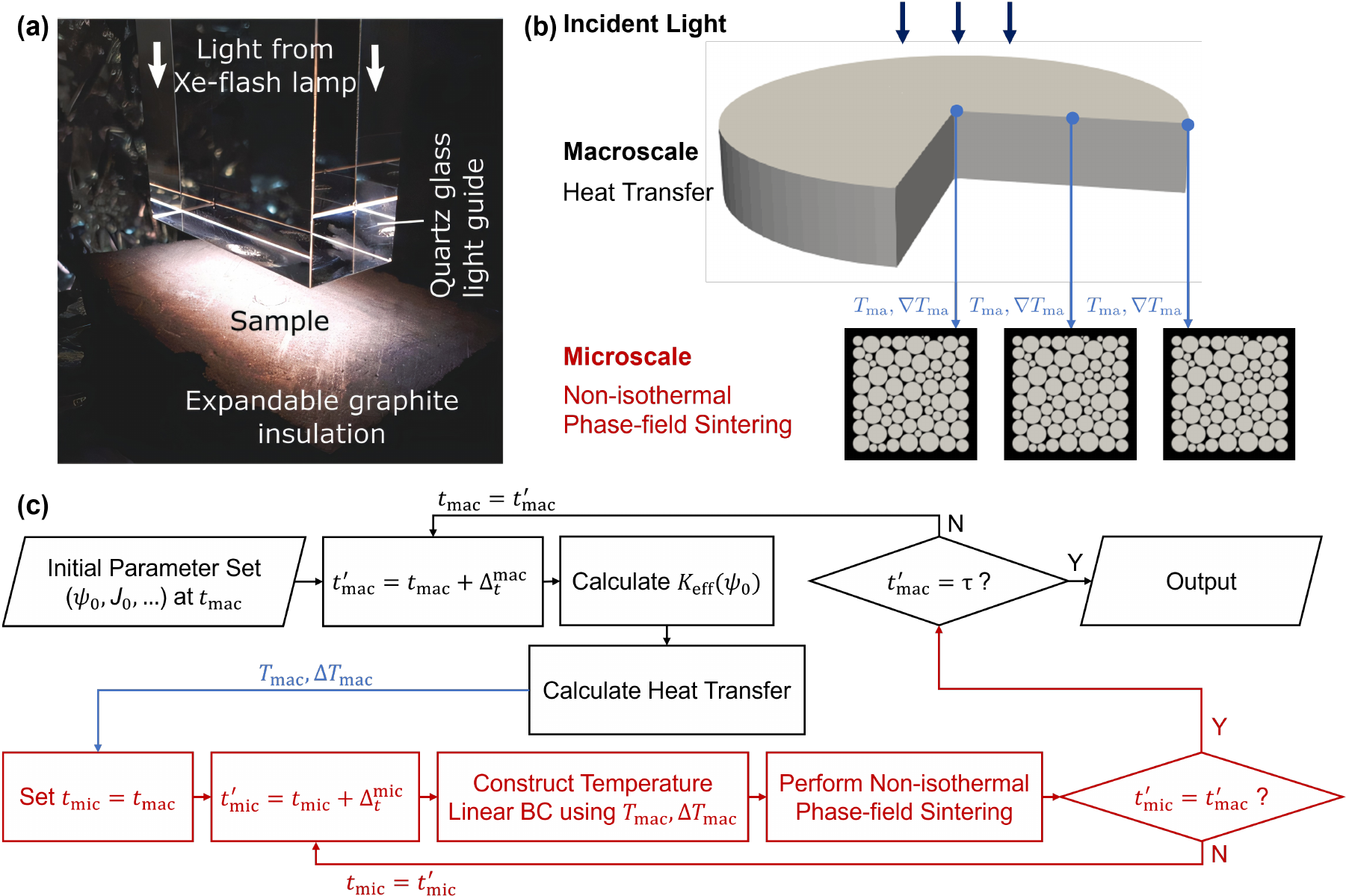}
    \caption{(a) Experimental setup for PS, image is reprinted with permission from Porz et al.\cite{porzBlacklightSinteringCeramics2022} under the terms of the Creative Commons Attribution 3.0 Unported Licence. (b) The schematic of the proposed two-scale simulation. (c) The workflow of the proposed two-scale scheme on every sampled point.}
    \label{fig:bs_wf_fc}
\end{figure}

Considering the thermal heterogeneity present within the sample, a fully resolved non-isothermal microstructural simulation, of the type commonly employed in additive manufacturing (AM), could in principle capture the relevant interactive effects, but at a considerably high computational cost. Notably, previous numerical studies of thermal profile during the photonic sintering process with a thermal insulating substrate have reported temperature gradients on the order of $0.1~\si{K~\micro m^{-1}}$ \cite{scherer2025sintering}, which are several orders of magnitude smaller than the $\sim 100~\si{K~\micro m^{-1}}$ gradients typically encountered in AM \cite{yang3DNonisothermalPhasefield2019}. In this regard, a staggered two-scale scheme is sufficient to accurately describe the coupled thermal–microstructural evolution, as shown in \subfigref{fig:bs_wf_fc}{b}.  On the macroscale, a three-dimensional (3D) FE-implemented heat transfer simulation is performed with the temperature $T$ as the sole degree-of-freedom (DoF). Owing to the sample's geometric symmetry and the assumption of a homogeneous initial packing density ($\psi_0$, which is also the initial volume fraction of substance), temperature gradients may develop in both the radial and thickness directions of the sample during PS process. Accordingly, sampling points are selected on a radial-thickness section of the macroscopic domain. 

To reduce computational cost while exploiting the cylindrical symmetry of the sample, microscopic non-isothermal phase-field sintering simulations are performed on a two-dimensional domain at each sampling point along a representative cross-section. Each phase-field simulation incorporates the conserved order parameter $\rho$ and a set of non-conserved order parameters $\{\phi_j\}$. In addition to the order parameters, the microscopic temperature field and the chemical potential are included as degrees of freedom, the latter arising from the split formulation of the Cahn--Hilliard equation \cite{yang3DNonisothermalPhasefield2019}.
Previous studies \cite{greenquist2020development, gong2019quantitative} have justified that both 2D and 3D simulations can capture the same underlying transport and coarsening mechanisms, despite existing quantitative differences. Consequently, the 2D sintering simulation provides a computationally efficient and physically consistent framework for coupling the microscale phase-field simulations with the macroscale cylindrical description.

The workflow of the proposed two-scale photonic sintering simulation is illustrated in \subfigref{fig:bs_wf_fc}{c}. At each time step ($t_\mac$), the macroscopic heat transfer problem (\eqref{gemacro}) is firstly solved using the effective thermal conductivity $K_{\mathrm{eff}}$, which is computed from the substance volume fraction $\psi$. This yields the macroscopic temperature field $T_\mac$ and its spatial gradient $\nabla T_\mac$. These local macroscopic quantities are then transferred to the microscopic domains and reconstructed as linear temperature BCs, ensuring satisfaction of the Hill-Mandel condition. Specifically, the consistency between the macroscopic and microscopic temperature fields and their gradients is enforced as
\begin{equation}
T_\mac = \langle T_\mic \rangle, \qquad 
\nabla T_\mac = \langle \nabla T_\mic \rangle,
\label{temp_homogen}
\end{equation}
where $\langle \cdot \rangle$ denotes the volumetric average, defined as $\langle \cdot \rangle = \Omega^{-1} \int_{\Omega} (\cdot)\,\diff \Omega$, with $\Omega$ representing the volume of the microscopic simulation domain. Subsequently, non-isothermal phase-field simulations (\eqref{gemesoconcentration}-\eqref{gemesoheat}) are carried out under the imposed BCs to resolve the thermally coupled microstructural evolution, thereby capturing the influence of the macroscopic thermal field on microscale sintering behavior. It should be noted that the time steps employed at the macroscale and microscale, denoted by $\Delta_t^\mac$ and $\Delta_t^\mic$, respectively, are not necessarily identical. Owing to numerical stability and convergence constraints (see Fig.~3 in Ref.~\cite{yang2025highthroughput}), the microscale simulations generally require smaller time steps, i.e., $\Delta_t^\mic \leq \Delta_t^\mac$ for most of time. Consequently, an inner iteration scheme is also implemented to synchronize the microscopic and macroscopic time integration during each macroscopic timestep, as also shown in \subfigref{fig:bs_wf_fc}{c}.

It is also worth noting that the effective thermal conductivity $K_{\mathrm{eff}}$ used in the macroscale simulation should, in principle, depend on the transient substance volume fraction $\psi(t)$. In this first two-scale modeling effort, however, only the effective thermal conductivity of the starting powder bed is considered. Several analytical homogenization models are employed to estimate the effective thermal conductivity of the initial powder assembly, as detailed in the following subsection.

\subsection{Macroscopic thermal evolution and photonic-ray–induced thermal effect}
\label{macrothermaltransfer}

Heat transfer in the macroscopic powder stack involves conduction through a sintered porous structure. Following the formulation proposed in Ref.~\cite{yang2022b_validated}, the governing equation for macroscopic heat transfer is expressed as
\begin{equation}
\label{gemacro}
    c_{\mathrm{eff}}(T)\,\pd{T}{t}
    = \nabla \cdot  K_{\mathrm{eff}}(T)\,\nabla T + q,
\end{equation}
where $c_{\mathrm{eff}}$ denotes the effective volumetric specific heat, $K_{\mathrm{eff}}$ is the effective thermal conductivity (assumed isotropic due to the random porous structure) and $q$ represents the photonic-ray–induced volumetric heat source. For porous materials, a variety of analytical homogenization models have been proposed to estimate $K_{\mathrm{eff}}$. These models primarily depend on the solid volume fraction $\psi$ and represent the porous medium as a composite in which a gaseous pore phase is embedded in a continuous ceramic matrix, with the underlying geometry idealized to varying degrees. In this work, five representative models summarized in Table~\ref{keff_model} are implemented and compared, where $K_{\mathrm{ss}}$ and $K_{\mathrm{at}}$ denote the thermal conductivities of the solid substance and the pore atmosphere, respectively.

\begin{table}
\centering
\renewcommand{\arraystretch}{1.5}
\caption{Analytical homogenization models for evaluation effective thermal conductivity of the powder stack, $\psi$ represents the volume fraction of the substance.}
\begin{tabular}{ccc}
\hline
\textbf{Model}                  & \textbf{Effective thermal conductivity equation}   & \textbf{Reference} \\ \hline
Reuss            & $K_{\text{eff}}=\frac{1}{\psi/K_{\text{ss}}+(1-\psi)/ K_{\text{at}}}$  &      \cite{reuss1929berechnung}     \\
Maxwell          & $K_{\text{eff}}= K_{\text{at}} \left(1 + \frac{3\psi}  {(K_{\text{ss}} + 2K_{\text{at}}) / (K_{\text{ss}} - K_{\text{at}}) - \psi}\right)$               &     \cite{maxwell1873treatise}      \\
{Zehner-Schlünder-Sih}             & $K_{\text{eff}}=K_{\text{at}}\left(\left(1-\sqrt{\psi}\right)+\sqrt{\psi}\left(\frac{2K}{1-BK_{\text{at}}/K_{\text{ss}}}\right)\right)^*$                                         &       \cite{sih1996}    \\
Reversed Maxwell & $K_{\text{eff}}= K_{\text{ss}} \left(1 + \frac{3 (1-\psi)} {(K_{\text{at}} + 2K_{\text{ss}}) / (K_{\text{at}} - K_{\text{ss}}) - (1-\psi)}\right)$ &           \\
Voigt           & $K_{\text{eff}}= \psi K_{\text{ss}} +(1-\psi) K_{\text{at}}$ &     \cite{voigt1889ueber}    \\ \hline
\end{tabular}
\captionsetup{justification=justified}
\caption*{\textbf{*:} In the ZSS model, 
\(
K = \frac{\left(1 - \frac{K_{\text{at}}}{K_{\text{ss}}} \right) B}
{\left(1 - \frac{B K_{\text{at}}}{K_{\text{ss}}} \right)^2}
\ln\left( \frac{K_{\text{at}}}{B K_{\text{ss}}} \right)
- \frac{B+1}{2}
- \frac{B-1}{1 - \frac{B K_{\text{at}}}{K_{\text{ss}}}},
\quad
B = 1.25 \left( \frac{\psi}{1 - \psi} \right)^{10/9}
\) 
}
\label{keff_model}
\end{table}

Accurate formulation of the photonic-ray–induced heat source is also essential for simulating the PS process. Notably, the incident beam energy is not confined to the surface of the powder stack but penetrates into it as photons propagate and scatter within particle gaps and voids, thereby generating a distributed thermal effect \cite{gusarov2009a}. Accordingly, the thermal effect of the incident photonic ray is modeled in this work as a volumetric heat source, whose depth-dependent distribution is described using a radiation penetration approach \cite{liu2018vheatsource, zhang2019vheatsource}. The powder bed is treated as a homogenized optical medium. The effective heat source $q$ at a coordinate $\hat{\mathbf{x}}$ within the macroscopic domain can be expressed as
\begin{equation}
    q[\hat{\mathbf{x}}(\mathbf{r}_\mathrm{S},z),t]=J_\mathrm{d}(\mathbf{r}_\mathrm{S}, t)p_z(z),
    \label{eq_q}
\end{equation}
where the coordinate of point $\hat{\mathbf{x}}$ is represented by the surface coordinate $\mathbf{r}_\mathrm{S}$ and thickness coordinate $z$. $J_\mathrm{d}$ is the nominal beam power density on the surface of the sample, which is assumed to follow a 2D uniform profile. $p_z(z)$ is the depth-dependent penetration function proposed by Gusarov et al. \cite{gusarov2009a}, which is formulated as 
\begin{equation}
    p_z(z)=-\eta(\lambda, D_{l})\overline{\beta}(d_{50},\varphi)\frac{\diff \bar{J}(A,\ell)}{\diff \ell},
\end{equation}
where $\bar{J}(A,\ell)$ is the dimensionless net radiative energy flux density within a close-packed powder stack, given by Gusarov \etal \cite{gusarov2009a}. $A$ is the hemispherical absorptivity of the powder, $\ell$ is the optical thickness. Additional factors such as light wavelength $\lambda$ and beam size that affect the attenuation of the power density on the surface are represented by an efficiency parameter $\eta (\lambda, D_l)$. This parameter is usually calibrated based on experimental measurements \cite{yang2022b_validated}. $\overline{\beta}$ represents the mean attenuation coefficient and it is formulated with the substance volume fraction $\psi$ and the median diameter of the powder $d_{50}$ as $\overline{\beta}=\frac{3}{2} \frac{\psi}{1 -\psi} \frac{1}{d_{50}}$. Given $\ell=\overline{\beta}z$, the formulation of \eqref{eq_q} can be rewritten as
\begin{equation}
    q = -\eta J_ \mathrm{d}\overline{\beta}\frac{\mathrm{d}\bar{J}}{\mathrm{d}\ell}.\label{eq:abs_profi}
\end{equation}
Depending on the processing sequence, containing heating, power-holding, and power-decay stages, $J_ \mathrm{d}$ can be prescribed as a time-dependent function, which is illustrated in the inset of \subfigref{fig:4_macro}{c}.

\subsection{Microscopic non-isothermal phase-field simulation of microstructure evolution}
\label{mesopfsimulation}

The non-isothermal phase-field model is employed to simulate the coupled thermal-structural evolution of the sample, following the work of Yang \textit{et al.} \cite{yang3DNonisothermalPhasefield2019, yang2020investigation}. In this model, a conserved order parameter (OP) $\rho$ is introduced to distinguish the substance phase ($\rho=1$) from the atmosphere/pore phase ($\rho=0$), and a set of non-conserved OPs $\{\phi_j\}$, with $j=1,2,\ldots$, is used to differentiate individual particles according to their crystallographic orientations. To ensure that the orientation fields $\{\phi_j\}$ are defined exclusively within the substance phase ($\rho=1$), a penalty constraint $\sum_j \phi_j + (1-\rho) = 1$ is imposed on the simulation domain, which is achieved by the penalty function method \cite{yang3DNonisothermalPhasefield2019}. The total free-energy density functional over the domain $\Omega$, which is temperature-dependent and expressed in terms of the above order parameters, can be written as
\begin{equation}
\mathcal{F} (T,\rho,\{ \phi_j \})=\int_\Omega \left[ f(T,\rho,\{\phi_j\})+\frac{1}{2}\underline{\kappa}_{\mathrm{sf}}(T) \left|\nabla\rho\right|^2 + \frac{1}{2}\underline{\kappa}_{\mathrm{gb}}(T) \sum_j\left|\nabla\phi_j\right|^2 \right]\mathrm d\Omega .
\end{equation}
Here, $\underline{\kappa}_{\mathrm{sf}}$ and $\underline{\kappa}_{\mathrm{gb}}$ are gradient energy coefficients related to $\rho$ and \{$\phi_j$\}. The local free energy density $f$ is formulated in the form of a Landau-type polynomial as:
\begin{equation}
\begin{aligned}
     f\left(T, \rho,\left\{\phi_j\right\}\right)=&f_{\mathrm{ht}}(T,\rho,\left\{\phi_j\right\})+ \underline{W}_{\mathrm{sf}}(T)\left[\rho(1-\rho)^2\right]\\&+\underline{W}_{\mathrm{gb}}(T)\left\{\rho^2+6(1-\rho) \sum_i \phi_j^2-4(2-\rho) \sum_i \phi_j^3+3\left[\sum_i \phi_j^2\right]^2\right\},
\end{aligned}
\end{equation}
where $\underline{W}_{\mathrm{sf}} (T)$ and $\underline{W}_{\mathrm{gb}} (T)$ are the temperature-dependent barrier heights. The local free energy density $f\left(T, \rho,\left\{\phi_j\right\}\right)$ exhibits local minima in various regions, indicating the thermodynamic stability of the corresponding phases, such as substance, atmosphere/pores, and solid grains with different orientations. The presence of a temperature gradient, represented by the heat term ($f_{\mathrm{ht}}$), can further alter this stability by generating driving forces that promote local diffusion, grain coalescence, and thermal transfer \cite{yang2020investigation}. The governing equation of the conserved OP is given by:
\begin{equation}
    \frac{\partial\rho}{\partial t}=\nabla\cdot\left[\mathbf{M}_{\rho}(\rho,T)\cdot\nabla\mu-\mu\mathbf{M}_{\rho}^{\mathrm{th}}(\rho,T)\cdot\frac{\nabla T}{T}\right],
    \label{gemesoconcentration}
\end{equation}
where the term $\mathbf{M}_{\rho}\cdot\nabla\mu$ represents the flux driven by Fickian diffusion, while $\mu\,\mathbf{M}_{\rho}^{\mathrm{th}}\cdot\nabla T/T$ corresponds to the flux induced by thermophoresis (also known as the Soret effect). The reciprocal phenomenon associated with thermophoresis, commonly referred to as the Dufour effect, is neglected as it is negligible in highly condensed systems \cite{yang2020investigation}. By comparing the governing equation with Fick’s law and the theory proposed by Schottky \cite{schottky1965theory, zhang2012phase}, the rank-two diffusive mobility tensors $\mathbf{M}_{\rho}$ and thermophoresis mobility tensor $\mathbf{M}_{\rho}^{\mathrm{th}}$ are parameterized from the diffusivity tensor $\mathbf{D}$ as $\mathbf{M}_{\rho}=\mathbf{D}/\underline{F}$ and $\mathbf{M}_{\rho}^{\mathrm{th}}=Q_{\mathrm{th}}^{v}\mathbf{D}/(\mathcal{R}T\underline{F})$, respectively, where $\mathcal{R}$ is the gas constant, $\underline{F}$ is the characteristic volumetric energy, and $Q_{\mathrm{th}}^{v}$ denotes the transport heat of a vacancy. It is worth noting that the diffusivity tensor $\Dten$ accounts for the path-dependent nature of mass transport by incorporating multiple temperature-dependent diffusivities, associated with distinct diffusion pathways. These include diffusion through the substance ($D_\subs$), surface diffusion ($D_\sf$), and grain-boundary diffusion ($D_\gb$). The formulation and numerical implementation of this path-dependent diffusivity are detailed in Refs.~\cite{yang3DNonisothermalPhasefield2019, yang2023tailoring, yang2025highthroughput}.

Under the non-isothermal condition, the chemical potential $\mu$ is derived as
\begin{equation}
    \mu = \frac{\partial f}{\partial\rho}-\pd{\underline{\kappa}_{\mathrm{sf}}}{T}\left(T\nabla^2\rho+\nabla T\cdot\nabla\rho\right).
    \label{gemesochem}
\end{equation}
Despite its role in substance transport, $\nabla T$ also serves as an additional driving force in GB migration. The evolution of grain growth is governed by
\begin{equation}
    \frac{\partial\phi_{j}}{\partial t}= -M_{\phi}^{\mathrm{gb}}\left[\frac{\partial f}{\partial\phi_j}-\pd{\underline{\kappa}_{\mathrm{gb}}}{T}\left(T\nabla^2\phi_j+\nabla T\cdot\nabla\phi_j\right)\right],
    \label{gemesoorientation}
\end{equation}
where $M_{\phi}^{\mathrm{gb}}$ is the mobility, related to the GB migration mobility $G$ through $M_{\phi}^{\mathrm{gb}}=G\gamma_{\mathrm{gb}}/T\underline{\kappa}_{\mathrm{gb}}$ \cite{moelans2008}.

Last but not least, the heat transfer on microscale is governed by 
\begin{equation}
    c_{\mathrm{r}}(\rho,T)\frac{\partial T}{\partial t}
  =\nabla\cdot\mathbf{K}(\rho,T)\cdot\nabla T.
  \label{gemesoheat}
\end{equation} 
Here $c_{\mathrm{r}}$ is the relative specific heat, which is introduced to represent the difference in $c$ between the substance and the surrounding atmosphere. In this work, phase-dependent thermal properties, including $c_\mathrm{r}$ and $\Kten$, are interpolated from temperature-dependent properties of the pure phases. Specifically, the substance phase ($\rho=1$) is characterized by $c_{\mathrm{ss}}$ and $K_{\mathrm{ss}}$, while the atmosphere ($\rho=0$) is described by $c_{\mathrm{at}}$ and $K_{\mathrm{at}}$. The interpolation procedure and its implementation are detailed in Refs.~\cite{yang3DNonisothermalPhasefield2019, yang2023tailoring, yang2025highthroughput}.

Although thermophoresis has been reported in several condensed material systems and the corresponding transport heat $Q_{\mathrm{th}}^{v}$ has been quantified for ceramics, like yttria-stabilized zirconia ($Q_{\mathrm{th}}^{v}=511~\si{kJ/mol}$) \cite{biesuz2019microstructural,yang2020investigation}, its role in BZY protonic ceramics has not yet been established, and reliable material parameters are currently unavailable. Moreover, the thermal profile of photonic sintering differs substantially from those encountered in processes where thermophoresis is known to be significant. 
As discussed above, temperature gradients in photonic sintering are several orders of magnitude smaller than those in AM or field-assisted sintering processes ($\sim 0.1~\si{K/\micro m}$ vs. $\sim 100~\si{K/\micro m}$). Under such conditions, the contribution of thermophoresis effect is expected to be negligible compared to Fickian diffusion. Accordingly, as an initial two-scale modeling effort for photonic sintering, the thermophoresis mobility in BZY is set to $\|\mathbf{M}_{\rho}^{\mathrm{th}}\|=0$. This assumption effectively suppresses thermophoresis contributions in the microscopic non-isothermal sintering simulations, thereby allowing the dominant diffusion-driven mechanisms to be isolated. The influence of thermophoresis under extreme thermal gradients, as well as the identification of $Q_{\mathrm{th}}^{v}$ for BZY, will be addressed in future studies.

\section{Simulation setup}

\subsection{Simulation domains and boundary conditions}

\begin{figure}
    \centering
    \includegraphics[width=1\linewidth]{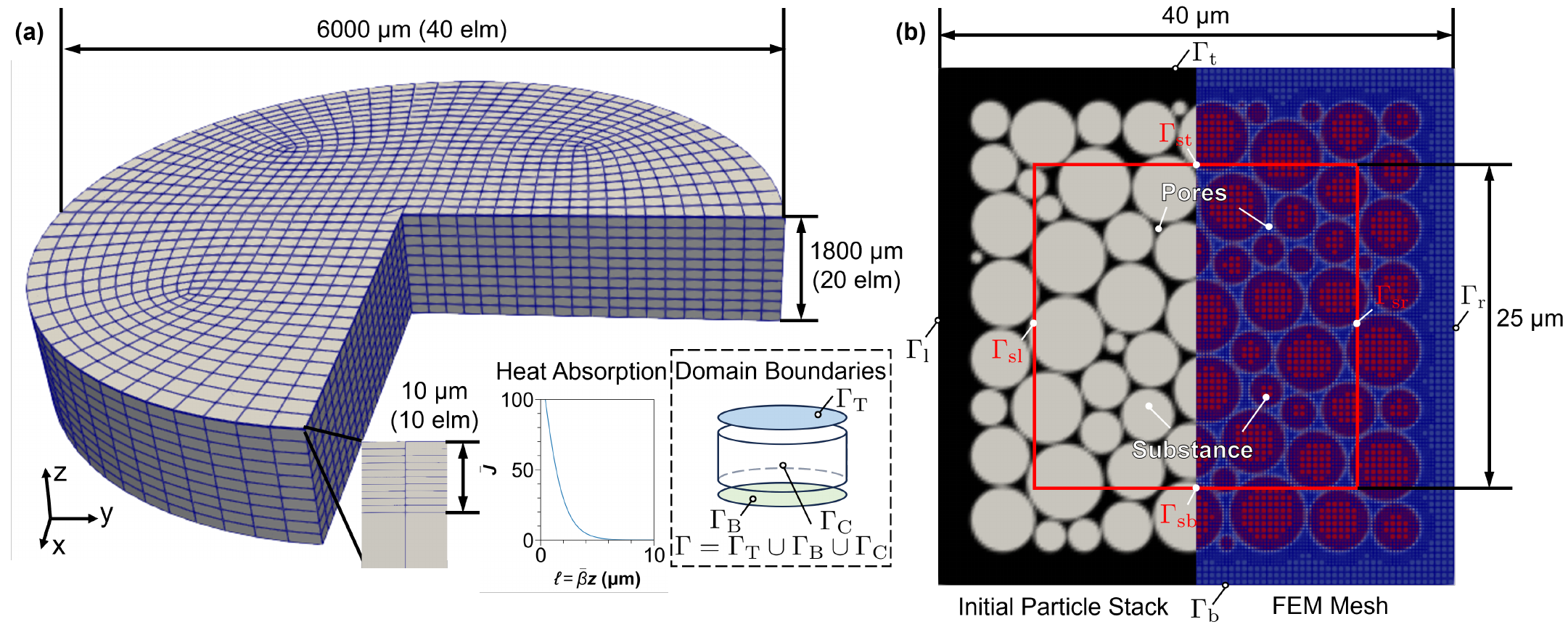}
    \caption{(a) Macroscopic mesh configuration showing element size and distribution, with insets illustrating the heat absorption profile and domain boundaries. (b) Microscopic initial powder stack with corresponding local mesh refinement.}
    \label{fig:ts_mesh}
\end{figure}

The finite-element mesh of the macroscopic simulation domain is shown in \subfigref{fig:ts_mesh}{a}. The mesh consists of twenty elements in the radial direction (corresponding to forty elements across the diameter) and twenty elements in the thickness direction. Mesh refinement is applied near the irradiated surface within a depth of $10~\si{\micro m}$ to adequately resolve the optical absorption profile, i.e., the $\mathrm{d}\bar{J}/\mathrm{d}\ell$ term in \eqref{eq:abs_profi}, as illustrated in the insets of \subfigref{fig:ts_mesh}{a}. Together with the governing equation \eqref{gemacro}, the following boundary conditions are imposed in the process simulations
\begin{equation}
    \label{macroBC}
    \left\{\begin{array}{l}\left.\nabla \rho\right|_{\Gamma} \cdot \hat{\mathbf{n}}=0 ,\\ \left.\mathbf{K} \cdot \nabla T\right|_{\Gamma_{\text {C }}\cup\Gamma_{\text {B}}} \cdot \hat{\mathbf{n}}=h_\at \left[\left(\left.T\right|_{\Gamma_{\text {C }}}-T_\mathrm{e}\right)+\epsilon_h \sigma_\mathrm{B}\left(\left.T\right|_{\Gamma_{\text {C }}} ^4-T_\mathrm{e}^4\right)\right], \\ \left.\frac{\partial T}{\partial n}\right|_{\Gamma_{\text {B }}}=0,\end{array}\right. 
\end{equation}
where $h_\at$ is the convectivity, $\sigma_\mathrm{B}$ is the Stefan-Boltzmann constant, $\epsilon_h$ is the hemispherical emissivity and $T_\mathrm{e}$ is the environmental temperature. The simulation boundaries and their labels are summarized in the inset of \subfigref{fig:ts_mesh}{a} with $\hat{\mathbf{n}}$ denoting the outward unit normal on each respective boundary.

The microscopic simulation domain, containing a close-packed powder stack with a mean particle radius of $2.050~\si{\micro m}$, is shown in \subfigref{fig:ts_mesh}{b}. The full domain has a dimension of $40 \times 40~\si{\micro m}^2$ and contains a highlighted sub-domain of size $25 \times 25~\si{\micro m^2}$, denoted as $\Gamma_{\mathrm{sub}}$. Within $\Gamma_{\mathrm{sub}}$, the macroscopic temperature field is imposed through linear temperature boundary conditions, consistent with the two-scale thermal homogenization scheme described in \eqref{temp_homogen}. 
Assuming a decoupling between the macroscopic and microscopic scales, mass conservation is enforced locally at the microscale by prescribing a zero-flux boundary condition for the density field $\rho$ on the outer boundary $\Gamma_{\mathrm{out}}$. Under these assumptions, the microscopic simulations are carried out using the following boundary conditions
\begin{equation}
    \left\{
    \begin{aligned}
      &  \left.\nabla \rho \right|_{\Gamma_{\text{out}}} \cdot \hat{\mathbf{n}} = 0, \\
      &  \left. T \right|_{\Gamma_{\text{sub}}} = T_\mac + \nabla T_\mac \cdot \hat{\mathbf{x}},
    \end{aligned}
    \right.
\end{equation}
where $T_\mac$ and $\nabla T_\mac$ denote the temperature and its gradient transferred from sampling points on the macroscopic cylindrical domain, $\hat{\mathbf{x}}$ represents the spatial coordinates on the boundaries, $\Gamma_{\text{out}} = \Gamma_\text{t} \cup \Gamma_\text{b} \cup \Gamma_\text{l} \cup \Gamma_\text{r}$, and $\Gamma_{\text{sub}} = \Gamma_\text{st} \cup \Gamma_\text{sb} \cup \Gamma_\text{sl} \cup \Gamma_\text{sr}$, with the boundary labels illustrated in \subfigref{fig:ts_mesh}{b}. Due to the usage of $h$-adaptive mesh (\subfigref{fig:ts_mesh}{b}), the initial static FE mesh of $80 \times 80$ quadrilateral elements is set for the microscopic domain. To capture evolving microstructural features with high fidelity, mesh refinement is adaptively applied based on the OPs ($\rho$ and ${\phi_j}$). The smallest element size reaches approximately $0.25~\si{\micro m}$ near particle interfaces. 

In this work, the degree of densification within the sub-domain is defined as
\begin{equation}
    \dens \equiv \frac{\int_{\Omega_{\mathrm{sub}}} \rho \, \diff \Omega}{\Omega_{\mathrm{sub}}},
\end{equation}
where $\Omega_{\mathrm{sub}}$ denotes the area of the sub-domain. To ensure complete densification ($\dens = 100\%$) at the grain-growth stage of sintering, when nearly all closed pores are eliminated, a buffer region is intentionally introduced between the close-packed particle assembly and the outer boundary ($\Gamma_\mathrm{out}$) of the microscopic domain. This buffer region accommodates geometric shrinkage during densification and prevents the formation of artificial porosity induced by shape variations of the powder stack. By evaluating densification solely within the designated sub-domain, those artificial effects are effectively excluded from the densification assessment.

\subsection{Parameterization and calibration}

\begin{table}
\centering
\renewcommand{\arraystretch}{1.5}
\caption{Material properties used in simulations.}
\begin{tabular}{cccc}
\hline
\textbf{Properties} & \textbf{Expressions ($T$ in K)} & \textbf{Units} & \textbf{References} \\
\hline
$T_0$           & $2273$ & $\mathrm{K}$ & \\
$T_\text{s}$           & $1873$ & $\mathrm{K}$ & \\
$\gamma_\text{sf}$      & $1$ & $\mathrm{J}/\mathrm{m}^2$ & \cite{rahaman_2003}\\
$\gamma_{\text{gb}}$    & $0.2$ & $\mathrm{J}/\mathrm{m}^2$ & \cite{rahaman_2003}\\
$D_{\text{sf}}$         & $2.38 \times 10^{-10} \exp \left( \frac{-1.42 \times 10^5}{ \mathcal{R}T} \right)$ & $\mathrm{m}^2/\mathrm{s}$ & \\
$D_{\text{gb}}$         & $7.34 \times 10^{-12} \exp\left(\frac{-1.42 \times 10^5}{\mathcal{R}T}\right)$ & $\mathrm{m}^2/\mathrm{s}$ & \cite{heoc2024}\\
$D_{\text{ss}}$         & $4.24 \times 10^{-17} \exp\left(\frac{-7.29 \times 10^4}{\mathcal{R}T}\right)$ & $\mathrm{m}^2/\mathrm{s}$ & \cite{heoc2024}\\
$K_{\text{BZY}}$ & $2.8$ & $\mathrm{J}/(\mathrm{s\, m\, K})$ & \cite{liu2018}\\
$K_{\text{Ar}}$  & $0.06$ & $\mathrm{J}/(\mathrm{s\, m\, K})$ & \cite{hoshino1986}\\
$\text{c}_{\text{BZY}}$ & $3.153 \times 10^6$ & $\mathrm{J}/(\mathrm{m^3\, K})$ & \cite{liu2018}\\
$\text{c}_{\text{Ar}}$  & $717.6$ & $\mathrm{J}/(\mathrm{m^3\, K})$ & \cite{chase1998nist}\\
$\bar{d}$               & $2.879$ & $\si{\micro m}$ & \\
$\sigma$                & $3.9825$ & $\si{\micro m}$ & \\
\hline
\end{tabular}
\label{tab:material_properties}
\end{table}

The material properties used to parameterize the PS process simulations are summarized in \tabref{tab:material_properties}. Given the following equations \cite{yang3DNonisothermalPhasefield2019,yang2023tailoring}
    \begin{equation}
        \begin{aligned}
           \gamma_{\mathrm{sf}}& =\frac{\sqrt{2}}{6} \sqrt{\left(W_{\mathrm{sf}}+7 W_{\mathrm{gb}}\right)\left(\kappa_{\mathrm{sf}}+\kappa_{\mathrm{gb}}\right)}, \\
           \gamma_{\mathrm{gb}} & =\frac{2 \sqrt{3}}{3} \sqrt{W_{\mathrm{gb}} \kappa_{\mathrm{gb}}} ,\\
           \ell_{\mathrm{gb}} & \approx \frac{2 \sqrt{3}}{3} \sqrt{\frac{\kappa_{\mathrm{gb}}}{W_{\mathrm{gb}}}}, \\
           \frac{W_{\mathrm{sf}}+W_{\mathrm{gb}}}{\kappa_{\mathrm{sf}}} &=\frac{6 W_{\mathrm{gb}}}{\kappa_{\mathrm{gb}}}, 
        \end{aligned}
        \label{eq:param}
    \end{equation}
the phase-field model parameters $W_{\mathrm{gb}}$, $W_{\mathrm{sf}}$, $\kappa_{\mathrm{gb}}$ and $\kappa_{\mathrm{sf}}$ can be derived from given $\gamma_{\mathrm{sf}}$, $\gamma_{\mathrm{gb}}$ and $\ell_{\mathrm{gb}}$, listed in \tabref{tab:material_properties}. It should be noted that both surface diffusion and grain-boundary diffusion play essential roles in sintering \cite{german2014sintering}. However, the surface diffusivity $D_{\mathrm{sf}}$ for BZY ceramics has not been reported in the literature. Accordingly, $D_{\mathrm{sf}}$ is calibrated against experimentally-measured temporal densification data before conducting the process simulations. This calibration enables the model to more accurately reproduce the densification kinetics observed in BZY ceramics.

\begin{figure}
    \centering
    \includegraphics[width=1\linewidth]{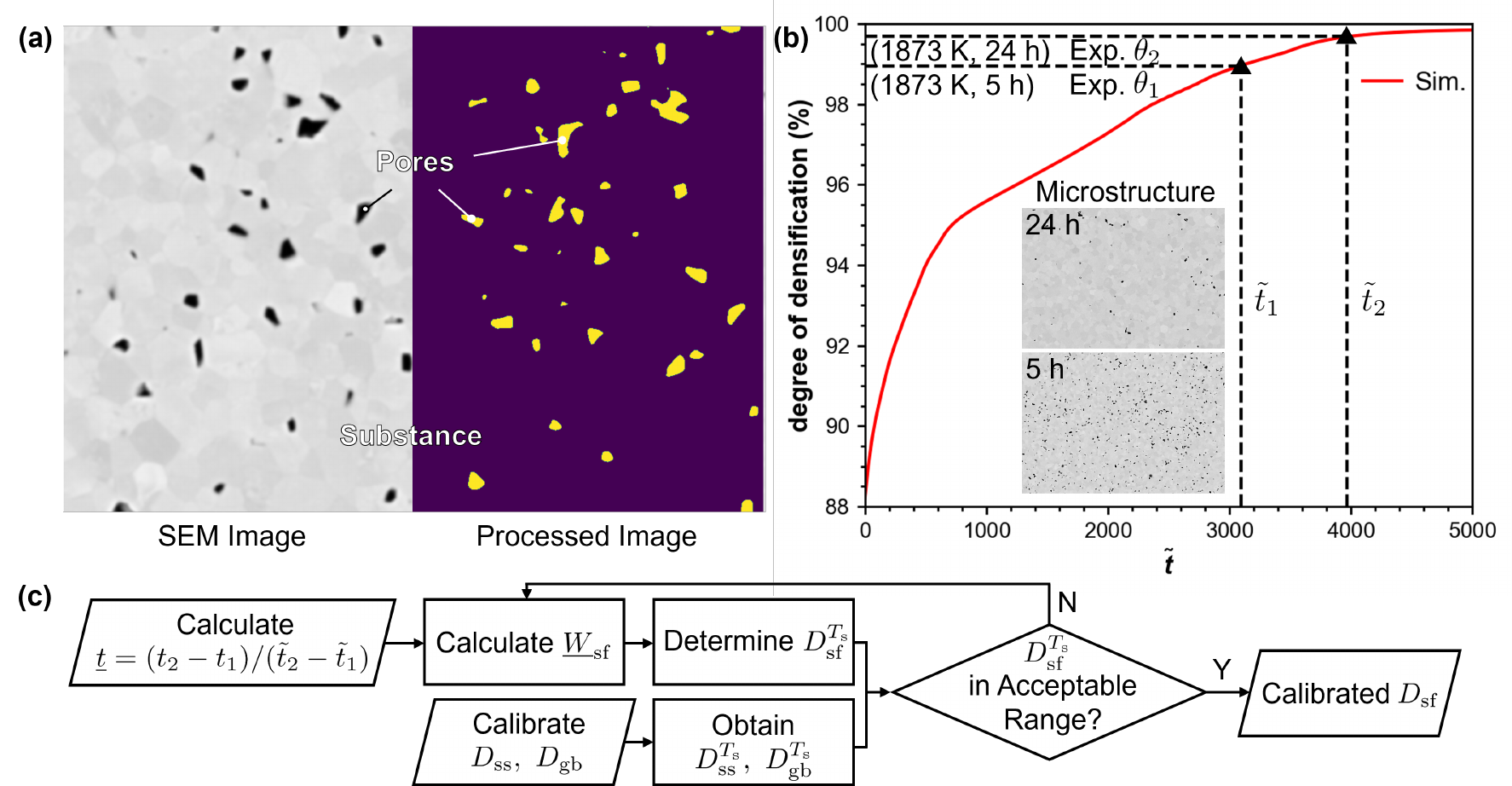}
    \caption{(a) Densification analysis: left - SEM image; right - thresholded image. (b) Isothermal densification curve at 1873 K, annotated with the degree of densification and corresponding times. Here the degree of densification $\dens$ is calculated from the porosity $\poro$ as $\dens=1-\poro$. (c) Workflow for calibrating the surface diffusivity. }
    \label{fig:cali_wf}
\end{figure}

We start with extracting temporal densification data from experimentally sintered BZY samples under controlled sintering temperature. These data are typically obtained from SEM images, as illustrated in \subfigref{fig:cali_wf}{a}. Following the image-processing workflow introduced in Ref.~\cite{mozhdeh2024adv}, each SEM image is preprocessed using a Gaussian blur to suppress noise and subsequently thresholded to segment pore regions. The resulting binary images enable quantitative evaluation of pore area and calculation of porosity and the corresponding densification degree $\dens$. SEM images of the same sample sintered for different durations and acquired at a magnification of $10~\si{\micro m}$ reveal a monotonic increase in densification with time, with $\dens_1 = 98.95\%$ after $t_{\mathrm{r}1} = 5~\mathrm{h}$ and $\dens_2 = 99.68\%$ after $t_{\mathrm{r}2} = 24~\mathrm{h}$. It is noted that grayscale similarity between certain pore regions and the matrix phase may occasionally lead to misclassification or under-segmentation, resulting in a slight overestimation of $\dens$. \subfigref{fig:cali_wf}{b} presents the calibrated simulation times, $\tilde{t}_1$ and $\tilde{t}_2$, along with representative SEM images, illustrating the microstructural evolution at the respective stages of densification. 

The workflow for calibrating $D_{\text{sf}}$ is illustrated in \subfigref{fig:cali_wf}{c}. First, the characteristic time is calculated. At a given sintering temperature, an isothermal phase-field sintering simulation is performed with solely an initial normalized value of $\tilde{M}^\sf_\rho=1$, which characterizes the modified diffusive mobility through the surface within the $\Mten_\rho$. Then, based on the simulated densification curve, the simulation time $\tilde{t}_1=3100.97$ and $\tilde{t}_2=3971.97$ are identified as corresponding to experimental densification degrees $\dens_1$ and $\dens_2$, respectively. The characteristic time is then established as by $\underline{t}=(t_\mathrm{r2}-t_\mathrm{r1})/(\tilde{t}_2-\tilde{t}_1)=78.53\ \text{s}$. Subsequently, counting the dimension of $M_\sf$ as $[\si{m^5~J^{-1}~t^{-1}}]$, the dimensional $M_\sf$ is calculated by $M_\sf=\tilde{M}^\sf_\rho \underline{l}^2 \underline{W}^{-1} \underline{t}^{-1}$ with the characteristic length $\underline{l}=1\ \si{\micro m}$, the characteristic energy density $\underline{W}=W_\text{sf}$ (with the dimension $[\si{J~m^{-3}}]$, obtained by \eqref{eq:param}), and the calibrated characteristic time $\underline{t}$. Finally, taking the correlation $M^\sf_\rho=D_\sf/\underline{F}$ (here $\underline{F}=2(W_\gb+W_\sf)$), the calibrated surface diffusivity can be calculated as
$D_{\mathrm{sf}}^{T_\mathrm{s}} = 2.59 \times 10^{-14}\ \mathrm{m}^2 /\mathrm{s}$ at the sintering temperature $T_\mathrm{s}$. This value is approximately two orders of magnitude higher than $D_{\mathrm{gb}}^{T_\mathrm{s}}=8.01\times 10^{-16}\ \mathrm{m}^2 /\mathrm{s}$ and five orders of magnitude higher than $D_{\mathrm{ss}}^{T_\mathrm{s}}=3.94 \times 10^{-19}\ \mathrm{m}^2 /\mathrm{s}$, an ordering within the margin of error relative to the experimental values \cite{heoc2024}. 

To consider the temperature dependency, $D_{\mathrm{sf}}$ is formulated by Arrhenius equation
\begin{equation}
\label{dsf}
    D_{\text{sf}}=D^{\text{sf}}_0 \mathrm{exp}\left(\frac{-Q_\mathrm{a}^\mathrm{sf}}{\mathcal{R}T}\right),
\end{equation}
where $D^{\mathrm{sf}}_{0}$ denotes the pre-exponential factor and $Q_{\mathrm{a}}^{\mathrm{sf}}$ is the activation energy for the surface diffusion. By assuming $Q_{\mathrm{a}}^{\mathrm{sf}}=Q_\mathrm{a}^\mathrm{gb}=1.42\times10^5$, $D^{\mathrm{sf}}_0$ can be calculated from $D_{\mathrm{ss}}^{T_\mathrm{s}}$ with $T=T_\mathrm{s}=1873$ \si{K}, yielding $D_\mathrm{sf} = 2.38 \times 10^{-10} \mathrm{exp} \left( -1.42 \times 10^5/ (\mathcal{R}T) \right)$, as listed in \tabref{tab:material_properties}.

\section{Results and Discussion}

\subsection{Thermal profile}

\begin{figure} [!h]
    \centering
    \includegraphics[width=1\linewidth]{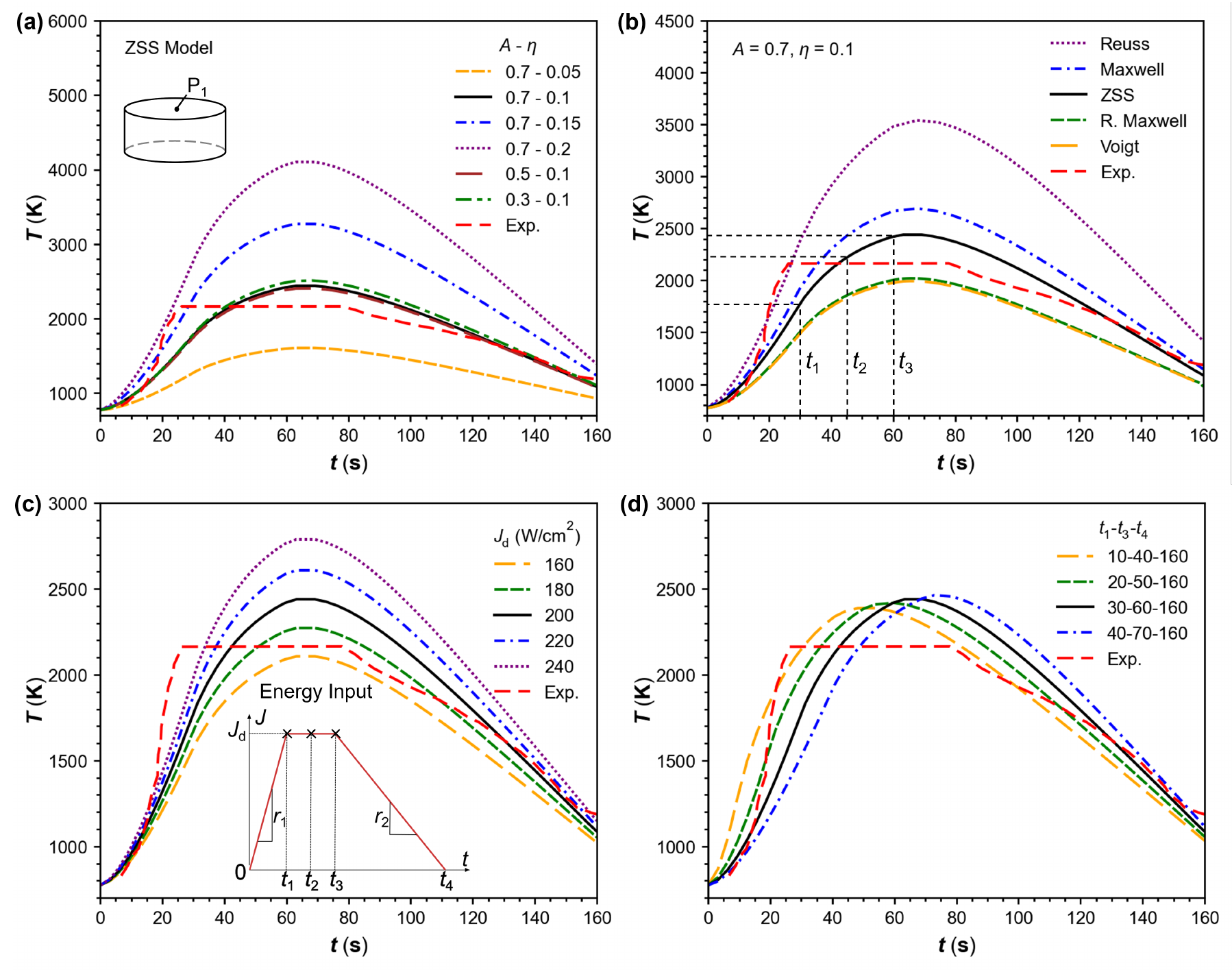}
    \caption{(a) Thermal history at the top central point $\mathrm{P}_1$ for varying powder absorptivity and light transfer efficiency, based on the ZSS model. Inset: schematic showing the location of $\mathrm{P}1$. (b) Thermal history at point $\mathrm{P}_1$ using different analytical effective thermal conductivity ($K_\text{eff}$) models, assuming constant porosity and fixed parameters ($A = 0.7$, $\eta = 0.1$). (c) Thermal history at point $\mathrm{P}_1$ under different maximum incident power flux $J_\mathrm{d}$. Inset: energy input profile, where $r_1$ and $r_2$ represent the power increasing and decreasing rates, respectively. (d) Thermal history at point $\mathrm{P}_1$ under various power loading paths.}
    \label{fig:4_macro}
\end{figure}

To investigate microstructure evolution during the PS process, it is essential to first determine the macroscopic thermal history. The location of the analyzed point, $\mathrm{P}_1$, is indicated in the inset of \subfigref{fig:4_macro}{a}. The incident power flux $J_\mathrm{d}(t)$ follows the programmed laser profile described in Sec.~\ref{methods}, with a maximum flux of $J_\mathrm{d}=200~\mathrm{W}/\mathrm{cm}^2$ (see inset of \subfigref{fig:4_macro}{c}). \subfigref{fig:4_macro}{a} shows the influence of powder absorptivity ($A$) and light transfer efficiency ($\eta$) on the resulting temperature evolution. The $K_\text{eff}$ is modeled using the ZSS model, as listed in Table~\ref{keff_model}, which concurrently captures contact and radiative effects in powders of diverse geometries~\cite{sih1996}. The powder absorptivity is defined as $A = 1 - \rho_\mathrm{h}$, where $\rho_\mathrm{h}$ denotes the hemispherical reflectivity of the powder. According to Refs.~\cite{ackermann2017a, vorobev2007}, typical reflectivity values range from 0.2 to 0.7, corresponding to $A$ values between 0.3 and 0.8.

Simulation results demonstrate that variations in $\eta$ have a more significant impact on the temperature field than those in $A$. Among the parameter sets examined, the combination of $A= 0.7$ and $\eta = 0.1$ shows the best agreement with experimental temperature profiles, reaching peak temperature of 2439~K. As shown in \subfigref{fig:4_macro}{a}, the temperature for this case increases from $t_1$ to $t_3$, reaching a peak around $t_3$. At the temperature level corresponding to the plateau in the measured data, the width of the simulated temperature profile closely matches that of the measured plateau. To further assess the influence of absorptivity, additional simulations were performed with a fixed $\eta = 0.1$ and varying $A$ in the range 0.3--0.7. The results reveal that, under low-efficiency conditions ($\eta = 0.1$), the system is relatively insensitive to changes in $\alpha$. Based on these observations and in accordance with reported optical properties of powder beds~\cite{ackermann2017a, vorobev2007}, the baseline parameter set $A = 0.7$ and $\eta = 0.1$ is adopted for subsequent simulations.

\begin{figure} 
    \centering
    \includegraphics[width=1\linewidth]{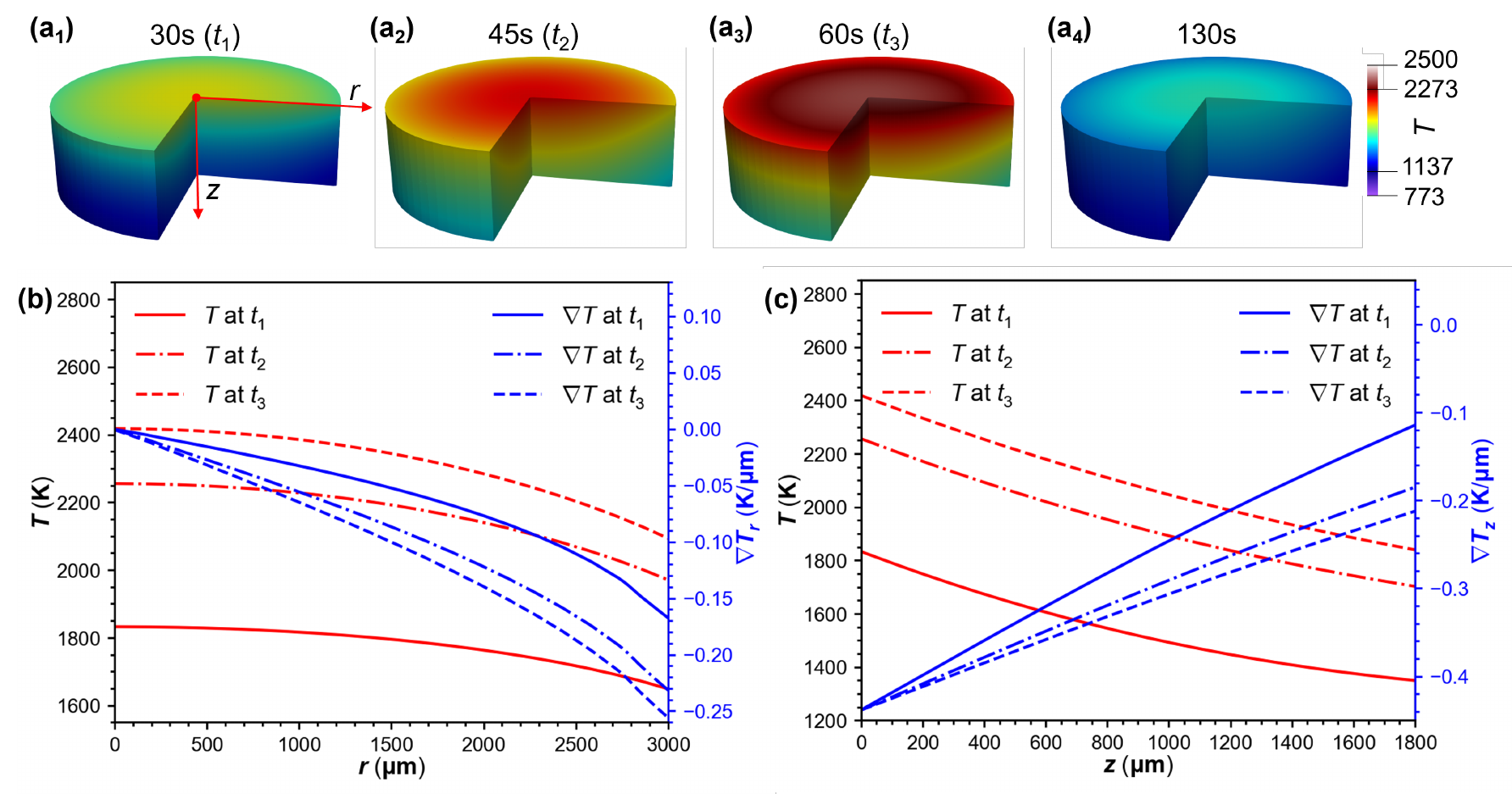}
    \caption{($\mathrm{a}_1$)–($\mathrm{a}_4$) Temperature distribution within the domain at different times. (b) Temperature and temperature gradient distributions along the radial direction ($r$). (c) Temperature and temperature gradient distributions along the thickness direction ($z$).}
    \label{fig:r1_spatial_TdT}
\end{figure}

\subfigref{fig:4_macro}{b} compares simulated temperature profiles obtained using various effective thermal conductivity models with experimentally measured data. All simulations begin at the same ambient temperature. However, as heating proceeds, the temperature trajectories diverge markedly, reflecting differences in the heat transport characteristics inherent to each model. The Reuss model predicts the fastest temperature rise, reaching a normalized peak value of approximately 3500~K near $t = 65$~s, indicative of limited thermal conductivity and significant heat accumulation. In contrast, the Voigt and reversed Maxwell models exhibit the slowest heating rates, with peak temperature below 2173~K, suggesting enhanced heat dissipation and higher effective thermal conductivities. The Maxwell and ZSS models show intermediate behavior, with peak values between 2173~K and 2700~K, indicating moderate thermal transport performance.

These differences highlight the critical influence of the selected effective thermal conductivity model on the predicted thermal behavior of the powder bed. Rapid heating corresponds to low effective conductivity and poor heat dissipation, whereas slower heating implies more efficient thermal transport. The simulation results are consistent with the theoretical bounds defined by the Voigt (upper) and Reuss (lower) limits and align with previous findings reported by Yang et al.~\cite{yang2022b_validated}. In this study, the ZSS model is adopted for subsequent analyses due to its intermediate thermal behavior, which is considered more representative of realistic powder bed characteristics in PS processes.

The simulated thermal histories also illustrate how variations in energy input parameters influence the temperature evolution at point $\mathrm{P}_1$, offering critical insight into material behavior during sintering. As shown in \subfigref{fig:4_macro}{c}, both the peak temperature and heating rate respond sensitively to changes in the maximum incident power flux $J_\mathrm{d}$. Specifically, increasing $J_\mathrm{d}$ leads to higher peak temperatures and faster heating, whereas decreasing $J_\mathrm{d}$ results in a more gradual thermal response. This demonstrates the central role of input photonic-ray intensity in controlling early-stage thermal dynamics across a broader range of conditions. \subfigref{fig:4_macro}{d} explores the influence of different power loading paths. {Although variations in $r_1$ and $r_2$ produce noticeable differences in peak temperature and heating/cooling rates, none of the simulated curves reproduce the extended high-temperature plateau observed in the experimental data (red dashed line). This discrepancy is primarily attributed to limitations in the measurement equipment, such as its finite response time and thermal inertia, rather than the actual thermal response characteristics of the material itself.}

The thermal response of the sample body was numerically investigated at various times (30~s, 45~s, 60~s and 130~s), as illustrated in \subssfigref{fig:r1_spatial_TdT}{a1}{a4}. Consistent with a centrally applied heat source, the top surface center of the sample exhibited the maximum temperature across all analyzed time points. Temperature progressively decreased with increasing radial ($r$) and thickness ($z$) distances, reaching its minimum at the sample's bottom edge. Quantitative analyses of the temperature profiles along the radial and thickness directions are presented in \subsfigref{fig:r1_spatial_TdT}{b}{c}, respectively. Along the radial direction, the temperature distributions at $t_1$, $t_2$, and $t_3$ exhibit parabolic profiles, with temperature differences between the center and the edge reaching 184~K, 285~K, and 324~K, respectively. The corresponding radial temperature gradients ($\nabla T_r$) show substantial spatial variation, reflecting strong lateral thermal heterogeneity. The negative sign of $\nabla T_r$ confirms that temperature decreases with increasing radial distance. Notably, the magnitude $|\nabla T_r|$ increases toward the sample edge, indicating the steepest thermal gradient occurs at the periphery. This pronounced gradient suggests a region of intense heat flux, which is critical for understanding potential microstructural evolution near the boundary.

Along the thickness direction, temperature differences were even more pronounced, measuring 483~K, 553~K, and 578~K at $t_1$, $t_2$, and $t_3$, respectively. Despite these higher absolute differences, the increase over time is more evident in the radial direction, indicating enhanced heat dissipation in the thickness direction. The temperature profiles along the thickness direction also follow a parabolic shape. However, the thickness gradients ($\nabla T_z$) decrease in magnitude with depth, reaching a minimum at the bottom surface ($z = 1800~\si{\micro m}$). This behavior contrasts with the radial direction, where the gradient magnitude increases toward the boundary. These opposing trends emphasize anisotropic thermal transport, with stronger gradients near the top surface and more thermally uniform conditions deeper within the sample. Given the presence of significant and depth-dependent thermal gradients along $z$, it is worthwhile to investigate microstructural evolution along the thickness direction.

\subsection{Microstructure and porosity}

\begin{figure} [!h]
    \centering
    \includegraphics[width=1\linewidth]{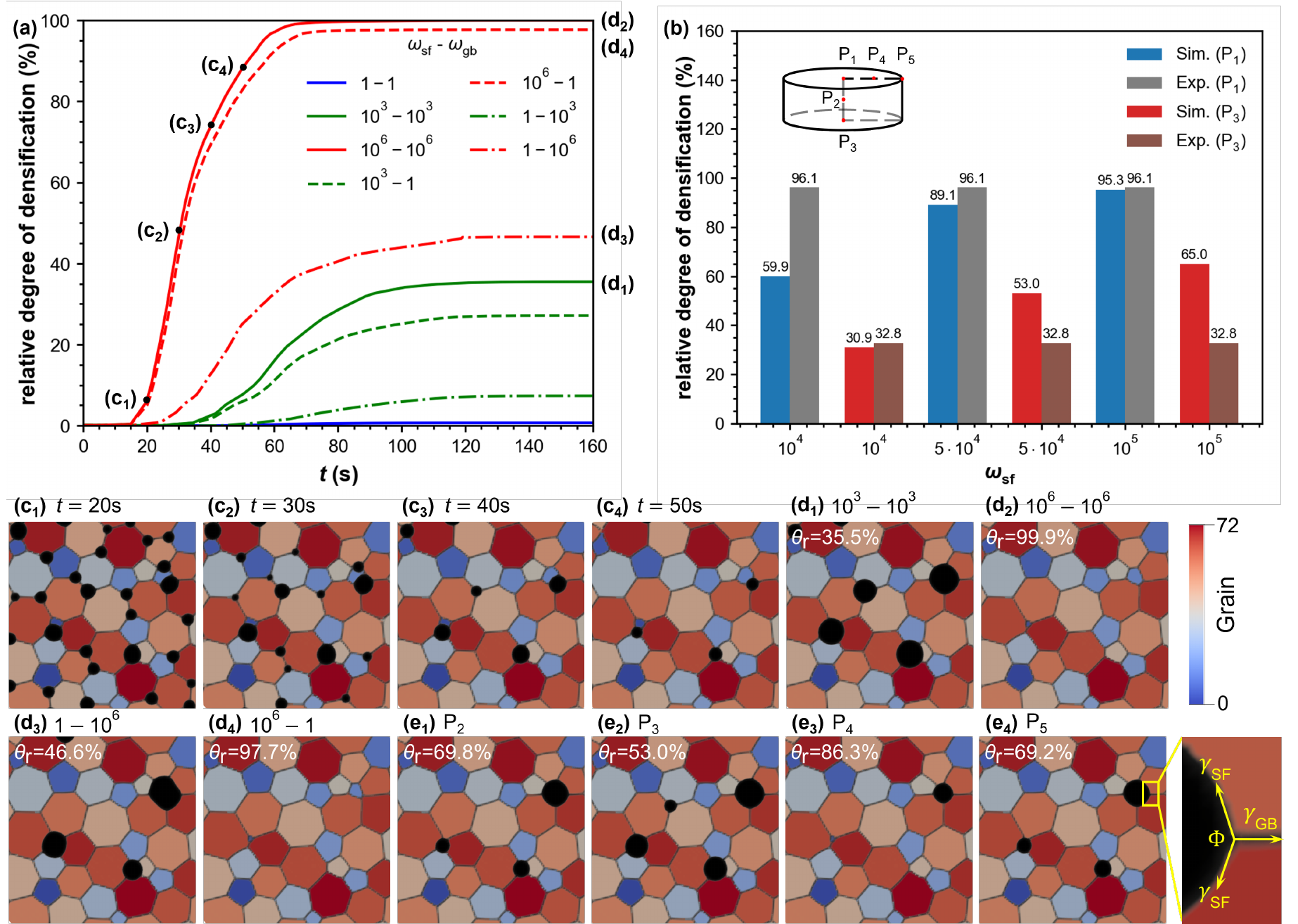}
    \caption{(a) Simulated relative degree of densification at $\mathrm{P}_1$ using enhanced mobility $M^\prime_\mathrm{sf}$ and $M^\prime_\mathrm{gb}$. (b) Comparison of relative degree of densification between simulated and experimental results at $\mathrm{P}_1$ and $\mathrm{P}_3$ using different $\omega_\mathrm{sf}$. ($\mathrm{c}_1$)-($\mathrm{c}_4$) Temporal microstructure evolution at fixed $\omega_\mathrm{sf}=10^6$ and $\omega_\mathrm{gb}=10^6$. ($\mathrm{d}_1$)-($\mathrm{d}_4$) Final microstructures obtained under different combinations of $\omega_\mathrm{sf}-\omega_\mathrm{gb}$. ($\mathrm{e}_1$)-($\mathrm{e}_4$) Final microstructures at $\mathrm{P}_2$–$\mathrm{P}_5$ with $\omega_\mathrm{sf}=5\times10^4$. Inset: Relationship between dihedral angle and surface/GB energy.}
    \label{fig:6_mgain}
\end{figure}

Rapid heating has been reported to suppress surface diffusion at low temperature, thereby preserving the driving force for densification \cite{rheinheimer2023impact}. Under steep thermal gradients, localized overheating at particle contacts can promote mass transport through softening or transient liquid phase triggered by thermal runaway \cite{biesuz2019microstructural, chaim2017particle}. Additionally, the suppression of coarsening and surface diffusion under such conditions has been linked to accelerated sintering kinetics via stress concentration and the formation of non-equilibrium porous structures \cite{maniere2023role}.

Simulations based on mobilities calibrated from isothermal sintering experiments fail to reproduce the experimentally observed densification of 95\% \cite{ebert2025a}, as shown by the blue line in \subfigref{fig:6_mgain}{a}.
This discrepancy suggests the emergence of additional thermally driven mechanisms beyond those described by conventional diffusion models. One plausible mechanism is the formation of a transient liquid phase at particle surfaces, which can substantially enhance mass transfer through surface and thereby accelerate densification under extreme thermal conditions. The emergence of such transient liquid phase may arise from multiple reasons, including chemical reactions between the powder and sintering aids, as well as partial melting associated with a locally reduced melting temperature induced by surface curvature or morphological effects. To account for this additional transport contribution, a dimensionless enhancement factor $\omega(T)$ is introduced in the model \cite{yang2023tailoring}. The calibrated mobility can be formulated as ($X=$sf, gb)
\begin{equation}
    M^\prime_X(T) = \omega_X(T) M_X(T).
\end{equation}

\subfigref{fig:6_mgain}{a} shows the simulated densification response as a function of the mobility enhancement factors $(\omega_\text{sf},\omega_\text{gb})$. The relative degree of densification ($\dens_\mathrm{r}$) is calculated as
\begin{equation}
    \dens_\mathrm{r}= \frac{\poro_0 - \poro}{\poro_0 - \poro_\mathrm{m}},
\end{equation}
where $\poro_0$ is the initial porosity and $\poro_\mathrm{m}$ is the minimum porosity that the microstructure can achieve. This $\poro_0$ can be related to the initial packing density $\psi_0$ as $\poro_0=1-\psi_0$ The baseline case $(1,1)$ yields negligible densification. When both mobilities are uniformly increased to $(10^3,10^3)$, the densification kinetics accelerate but saturate at only $\sim 35\%$, far below full densification. By contrast, $(10^6,10^6)$ produces a rapid sigmoidal trajectory, reaching nearly 100\% within $\sim 50$~s. The corresponding microstructure evolution is presented in \subssfigref{fig:6_mgain}{c1}{c4}, which clearly reveal the pore elimination process: isolated pores progressively shrink and vanish as particle contacts increase, accompanied by grain rearrangement and shape accommodation. This morphological evolution underlies the sharp sigmoidal densification trajectory observed for $(10^6,10^6)$.

Pore morphology and evolution during sintering are controlled primarily by the dihedral angle and particle coordination, which jointly determine the local equilibrium geometry and densification behavior. The relationship between the dihedral angle and surface/GB energy is given as $\gamma_\mathrm{gb}=2\gamma_\mathrm{sf}\mathrm{cos} ({\Phi}/{2})$, as illustrated aside of \subfigref{fig:6_mgain}{e4} As sintering proceeds, open pores with complex, interconnected shapes between grains gradually transform into isolated pores as the necks between them close. A larger dihedral angle indicates a lower ratio of GB energy to surface energy, which favors the formation of more spherical, energetically stable pores. Conversely, a smaller dihedral angle tends to promote the formation of elongated grooves extending along grain boundaries \cite{delannay2015sintering, kellett1989thermodynamics}. In this simulation, the surface energy was set to 1 $\mathrm{J}/\mathrm{m}^2$ and the grain-boundary energy to 0.2 $\mathrm{J}/\mathrm{m}^2$, yielding a dihedral angle of approximately 168.5$^\circ$. This very large angle implies that surface energy dominates, leading to nearly spherical equilibrium pores—consistent with the circular pore shapes observed in the results.

The asymmetric cases reveal a strong disparity between surface and grain-boundary contributions. When only surface mobility is enhanced $(10^6,1)$, the system follows nearly the same trajectory as the fully enhanced case $(10^6,10^6)$, achieving near-complete densification. In contrast, enhancing only grain-boundary mobility $(1,10^6)$ yields an intermediate plateau at $\sim 50\%$, demonstrating that grain-boundary diffusion alone is insufficient to eliminate porosity. Similar trends are observed at lower enhancement levels: $(10^3,1)$ achieves $\sim 25\%$ densification, whereas $(1,10^3)$ barely exceeds $5\%$. Representative final microstructures for different cases are shown in \subssfigref{fig:6_mgain}{d1}{d4}, which highlight the distinct pore morphologies: surface-dominated cases yield nearly pore-free packing, whereas grain-boundary-dominated cases preserve isolated or junction-trapped pores, consistent with their lower densification levels.

To identify the appropriate magnitude of $\omega_{\text{sf}}$, simulations were compared directly with experimental densification data at multiple positions within the sample (\subfigref{fig:6_mgain}{b}). The simulated relative degree of densification at $\mathrm{P}_1$ and $\mathrm{P}_3$ was systematically evaluated for $\omega_{\text{sf}}$ ranging from $10^4$ to $10^5$. Low enhancement factors ($\omega_{\text{sf}} \leq 10^4$) underestimate densification at $\mathrm{P}_1$, whereas values above $10^5$ overpredict densification at $\mathrm{P}_3$. Among the tested conditions, $\omega_{\text{sf}}=5\times10^4$ provides the better quantitative agreement with experimental measurements at both locations. The final microstructures at $\mathrm{P}_2$–$\mathrm{P}_5$ are presented in \subssfigref{fig:6_mgain}{e1}{e4}, where distinct porosity levels reveal the spatial heterogeneity of densification within the compact. Importantly, the radial variation in densification is relatively minor compared with the pronounced difference along the thickness direction, highlighting the anisotropic densification behavior across the compact. 

\begin{figure} 
    \centering
    \includegraphics[width=1\linewidth]{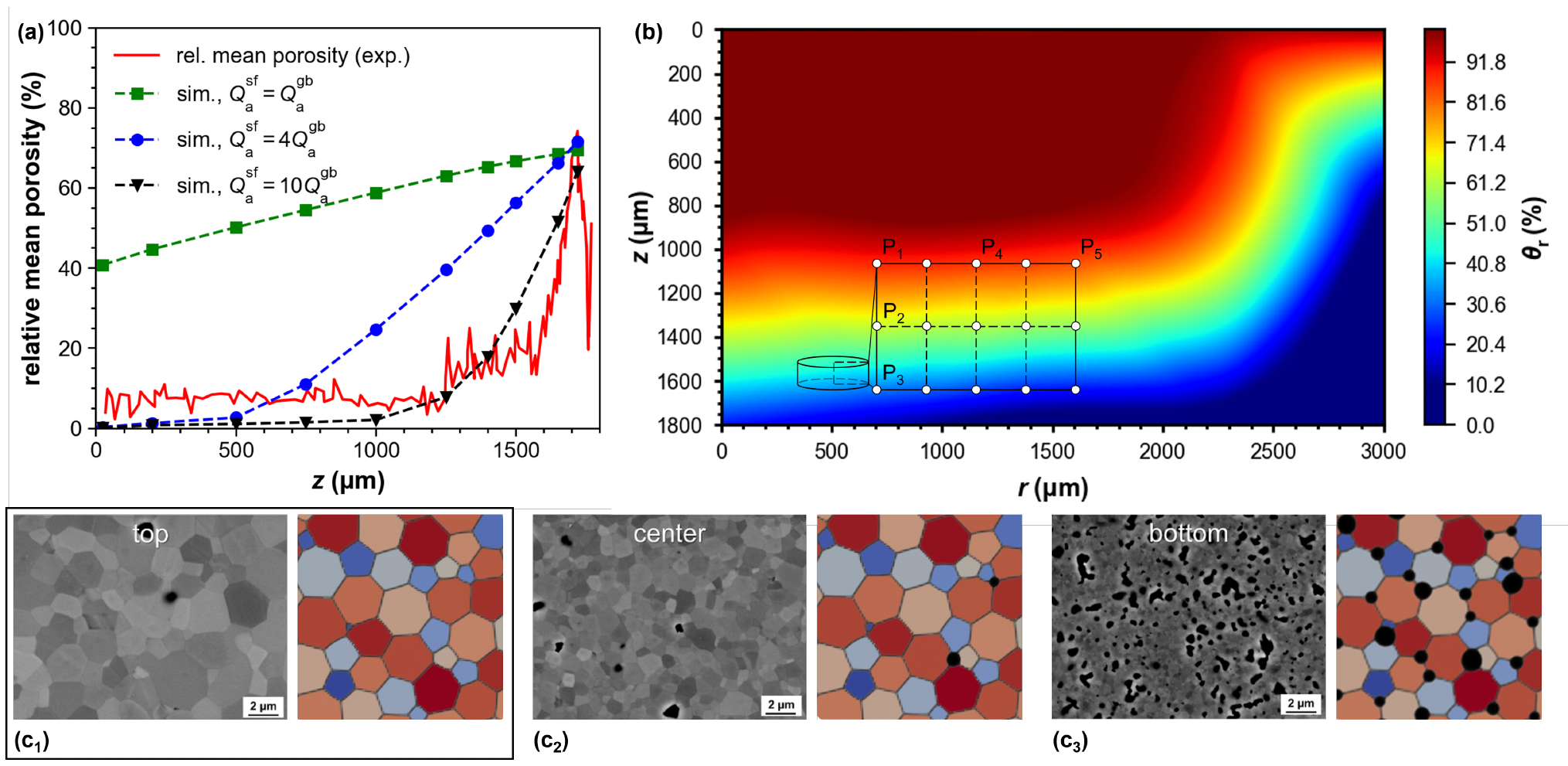}
    \caption{(a) Simulated relative mean porosity distribution along the thickness direction compared with experimental measurements. (b) Densification map of the half cross-section of the cylindrical sample implementing $Q_{\mathrm{a}}^\mathrm{sf} = 10 Q_{\mathrm{a}}^\mathrm{gb}$, the inset is the discrete sampling points used to calculate the relative degree of densification. (c$_1$)–(c$_3$) Microstructures at points P$_1$, P$_2$, and P$_{3}$ obtained using $Q_{\mathrm{a}}^\mathrm{sf} = 10 Q_{\mathrm{a}}^\mathrm{gb}$, shown alongside corresponding experimental SEM images.}
    \label{fig:7_overall}
\end{figure}

As shown in \figref{fig:6_mgain}, densification exhibits a clear positive correlation with surface mobility. Nevertheless, discrepancies arise when comparing simulation results with experimental measurements along this direction, as presented in \subfigref{fig:7_overall}{a}. While part of the deviation can be attributed to the underlying temperature distribution, a significant contribution originates from the spatial variation of mobility. When secondary effects associated with thermal gradients, such as the thermophoresis \cite{yang2020investigation}, are considered negligible, the dominant factor becomes the distribution of mobility along the depth direction. This distribution is strongly influenced by $Q_{\mathrm{a}}^\mathrm{sf}$, as shown in Eq.~(\ref{dsf}). Increasing $Q_{\mathrm{a}}^\mathrm{sf}$ enables a redistribution of mobility, thereby improving the consistency between simulated porosity evolution and experimental data.

The red fluctuating line in \subfigref{fig:7_overall}{a} represents the experimentally measured relative mean porosity as a function of depth. To reduce computational cost while still capturing the essential features of the porosity--depth relationship, ten representative points were selected from the experimental dataset for simulation ($z =$ 25, 200, 500, 750, 1000, 1250, 1400, 1500, 1650, 1720 \si{\micro m}). These points were chosen to provide full coverage of the investigated depth range, from near-surface to the deepest measurements, and to represent different porosity regimes. The spacing of selected depths was non-uniform: fewer points were chosen in regions where porosity varies slowly, while denser sampling was used in zones where porosity increases rapidly or nonlinear transitions are observed (e.g., between 1000--1720).

The simulated porosity values at the selected depths, obtained using same $\omega_{\text{sf}}=10^4$ and different values of $Q_{\mathrm{a}}^\mathrm{sf}$, are also shown in \subfigref{fig:7_overall}{a}. Increasing $Q_{\mathrm{a}}^\mathrm{sf}$ effectively shifts mobility towards deeper regions, leading to a stronger porosity growth with depth. Among the tested conditions, the case with $Q_{\mathrm{a}}^\mathrm{sf}=10Q_{\mathrm{a}}^\mathrm{gb}$ achieves the best agreement with experiment, reflected by the highest $R^2$ value (85.12\%). This suggests that using identical activation energies for surface and grain-boundary diffusion ($Q_{\mathrm{a}}^\mathrm{sf}=Q_{\mathrm{a}}^\mathrm{gb}$) underestimates the depth dependence of mobility, and that enhanced surface diffusivity at larger depths is required to capture the experimental porosity evolution. The remaining discrepancies, especially at the deepest regions where experimental porosity exhibits large fluctuations, are likely related to additional effects such as microstructural heterogeneity or transient thermal conditions not included in the present model.

The inset of \subfigref{fig:7_overall}{b} marks the discrete sampling points used to calculate the relative degree of densification ($\dens_\mathrm{r}$) and construct the contour map across half of the cross section. The resulting distribution shows a pronounced spatial gradient: $\dens_\mathrm{r}$ is highest near the top surface ($z \approx 0$) and progressively decreases with increasing depth. A radial variation is also visible, with densification decreasing from the center ($r=0$) towards the periphery. However, this effect is significantly weaker compared with the depth dependence. The final microstructures at P$_1$, P$_2$, and P$_{3}$ are presented in \subfigref{fig:7_overall}{c1}{c3} alongside the corresponding experimental SEM images. Notably, the simulated results exhibit less grain growth compared with the experiments. This indicates a further improvement of the model is needed.

\section{Conclusion}

In this work, a multiphysics two-scale simulation framework was developed to investigate the thermo-structural evolution of protonic ceramic during the photonic sintering (PS) process. 
\begin{itemize}
    \item Macroscopic heat transfer simulations revealed that optical absorption, thermal conductivity models, and power input strongly influence the macroscopic thermal history. Strong anisotropic temperature gradients, both radially and through the thickness, were identified as the primary drivers of heterogeneous densification.
    \item {At the microscopic scale, conventional diffusion-based mobilities were insufficient to simulate the densification mechanism in photonic sintering. Simulations indicate that enhanced surface diffusion, potentially induced by transient liquid phases at surface and/or grain boundary, should be incorporated to accurately reproduce the observed densification behavior.}
    \item By incorporating the calibrated surface activation energy, the experimentally observed pore gradient along the thickness direction was accurately predicted (with R$^2$=85.12\%), thereby establishing a clear process–microstructure relationship.
\end{itemize}
Overall, the developed framework demonstrates strong predictive capability for optimizing PS parameters and tailoring microstructural uniformity in protonic ceramics. Future work will extend the framework to a fully two-way coupled model, allowing feedback from evolving microstructure to modify the macroscopic thermal response. In-situ experimental validation will also be incorporated to further enhance predictive reliability.

\section*{Data Availability}
The authors declare that the data supporting the findings of this study are available within the paper. 

\section*{Code Availability}
Source codes of MOOSE-based application NIsoS and related utilities are available and can be accessed via the online repository \url{bitbucket.org/mfm_tuda/nisos.git}.

\section*{Funding Information}
    This study was funded by the German Science Foundation (DFG) under project 556363981 (FOR5966 SynDiPET - P3) and project 463184206 (CRC 1548, sub-projects B03). The funding agency played no role in study design, data collection, analysis and interpretation of data, or the writing of this manuscript.
    
	\section*{Acknowledgements}
    The authors greatly appreciate the access to the Lichtenberg II high-performance computer by the NHR Center NHR4CES@TUDa (funded by the German Federal Ministry of Education and Research and the Hessian Ministry of Science and Research, Art and Culture) and High-performance computer HoreKa by the NHR Center NHR@KIT (funded by the German Federal Ministry of Education and Research and the Ministry of Science, Research and the Arts of Baden-Württemberg, partly funded by the DFG). The NHR4CES Resource Allocation Board allocates computing time on the HPC under the project ``special00007''. 

\section{Competing Interests}
The authors declare no competing financial or non-financial interests.

\clearpage

\bibliography{reference, reference_Yang}
\end{document}